# Storage Coding for Wear Leveling in Flash Memories

Anxiao (Andrew) Jiang, *Member, IEEE,* Robert Mateescu, *Member, IEEE,* Eitan Yaakobi, *Student Member, IEEE,* Jehoshua Bruck, *Fellow, IEEE,* Paul H. Siegel, *Fellow, IEEE,* Alexander Vardy, *Fellow, IEEE,* and Jack K. Wolf, *Life Fellow, IEEE*

*Abstract*—Flash memory is a non-volatile computer memory comprised of *blocks of cells*, wherein each cell is implemented as either NAND or NOR floating gate. NAND flash is currently the most widely used type of flash memory. In a NAND flash memory, every block of cells consists of numerous *pages;* rewriting even a single page requires the whole block to be erased and reprogrammed. Block erasures determine both the longevity and the efficiency of a flash memory. Therefore, when data in a NAND flash memory are reorganized, minimizing the total number of block erasures required to achieve the desired data movement is an important goal. This leads to the *flash data movement problem* studied in this paper. We show that coding can significantly reduce the number of block erasures required for data movement, and present several optimal or nearly optimal *data-movement algorithms* based upon ideas from coding theory and combinatorics. In particular, we show that the sorting-based (non-coding) schemes require at least $O(n \log n)$ erasures to move data among $n$ blocks, whereas coding-based schemes require only $O(n)$ erasures. Furthermore, coding-based schemes use only one auxiliary block, which is the best possible, and achieve a good balance between the number of erasures in each of the $n + 1$ blocks.

*Index Terms*—coding theory, combinatorics, data storage, flash memory, graph theory, wear leveling

## I. INTRODUCTION

FLASH memory devices have become the most widely used non-volatile electronic memories. There are two basic types of such devices: NAND and NOR flash memories [8]. Between them, NAND flash is currently used much more often due to its higher data density. In a NAND flash, floating-gate cells are organized into *blocks*. Each block is further partitioned into multiple *pages*, and every read or write operation accesses a page as a unit. Typically, a page has 2KB to 4KB of data, and 64 or 128 pages comprise a block [8]. Flash memories have a unique *block erasure* property: although every page can be read and written (for the first time) individually, rewriting a page (that is, modifying its contents) requires the whole block to be erased and then reprogrammed. Typically, every block can endure $10^4 \sim 10^5$ erasures, after which the flash memory no longer meets quality guarantees and may break down. Block erasures also degrade the quality of the cells, introduce errors in the data, and reduce the overall read and write performance. Therefore, it is critical to minimize the number of block erasures. It is also critical to balance the number of erasures across different blocks. For this reason, numerous *wear leveling* techniques are widely used in flash-memory systems. The general idea is to balance erasures by migrating data to different locations, especially when data are rewritten [8].

In wear leveling, it is often desirable to move the frequently changing data (so-called hot data) into the same blocks, while storing the mostly static data together in other blocks. Thereby the overall erasures caused by the hot data can be reduced (see [8], [9]). The specific locations to which the data are moved can be optimized not only based on the update frequencies, but also on the correlation among the data. Another important application where data movement is required is *defragmentation of files*. Many file systems (and database systems) implemented in flash take the log-structured approach, wherein updates to files are stored non-consecutively across blocks. This way, wear leveling is achieved and local block erasures are avoided [5]. Consequently, files are frequently fragmented. To improve performance, data have to be moved periodically in order to reorganize the file segments. In database systems or sensors, after bursty incoming data flows are reliably stored, data movement is used to store the data in a categorized manner for efficient analysis. To facilitate data movement, a flash translation layer (FTL) is usually employed to map logical data pages to physical pages in the flash memory [8]. Minimizing the number of block erasures incurred during the data movement process remains a major challenge.

In this paper, we show that coding techniques can significantly reduce the number block erasures incurred during data movement. In addition to the overall number of erasures, we also consider other parameters, such as coding complexity and extra storage space (number of auxiliary blocks). We show that without coding, at least two auxiliary blocks are needed to enable data movement, and present a sorting-based solution that requires $O(n \log n)$ block erasures in order to move data among $n$ blocks. With coding, only one auxiliary block is needed. We present a very efficient data-movement algorithm based on coding over GF(2) that requires only $2n$ erasures.

Anxiao (Andrew) Jiang is with the Department of Computer Science and Engineering, Texas A&M University, College Station, TX 77843-3112, U.S.A. (e-mail: ajiang@cse.tamu.edu).

Robert Mateescu and Jehoshua Bruck are with the Department of Electrical Engineering, California Institute of Technology, 1200 East California Blvd., Pasadena, CA 91125, U.S.A. (e-mail: {mateescu,bruck}@caltech.edu).

Eitan Yaakobi, Paul H. Siegel, Alexander Vardy and Jack K. Wolf are with the Department of Electrical and Computer Engineering and the Center for Magnetic Recording Research, University of California at San Diego, La Jolla, CA 92093, U.S.A. (e-mail: {eyaakobi,psiegel,avardy,jwolf}@ucsd.edu).

This research was supported in part by the NSF CAREER Award CCF-0747415, NSF grants ECCS-0802107 and CCF-0830752, Caltech Lee Center for Advanced Networking, and the Center for Magnetic Recording Research at University of California, San Diego. Part of the results in the paper were presented at the IEEE International Symposium on Information Theory, Seoul, Korea, June 28 to July 3, 2009 (reference [12]).

We also present another coding-based algorithm that requires at most $2n-1$ erasures, which is worst-case optimal. Although minimizing the number of erasures for every instance of data movement is NP hard, as we prove in Section V-B, both algorithms achieve an approximation ratio of 2 with respect to the minimum possible number of erasures for the given instance.

We note that a number of papers on coding for flash memories have recently appeared in the literarture. These include codes for efficient rewriting [3], [7], [10], [15] (also known as *floating codes* or *flash codes*), error-correcting codes [4], and rank-modulation codes for reliable cell programming [11], [13]. However, to the best of our knowledge, this paper is the first to address storage coding at the page level instead of the cell level. Furthermore, our topic of study, namely the data-movement problem, is also distinct from all previous works.

The rest of the paper is organized as follows. In Section II, we define the data movement problem and introduce the relevant notation. In Section III, sorting-based data movement algorithms are presented. We further show in Section III that coding can help minimize the auxiliary storage requirements during data movement. In Section IV, we develop an efficient data-movement algorithm, based on coding over $GF(2)$. This algorithm uses only $2n$ erasures in order to move data among $n$ blocks. In Section V, we present an alternative coding-based algorithm, which uses at most $2n - 1$ erasures and is worst-case optimal. The NP hardness of the general problem of minimizing the number of erasures in data movement (for every given instance) is also established in V. Finally, Section VI contains our concluding remarks.

## II. Terms and Concepts

In this section, we formally define the data movement problem, and present some useful concepts.

**Definition 1.** (Data Movement Problem)
*Consider $n$ blocks storing data in a flash memory, and suppose that each block contains $m$ pages. The $n$ blocks are denoted by $B_1, \ldots, B_n$, and the $m$ pages in block $B_i$ are denoted by $p_{i,1}, \ldots, p_{i,m}$ for $i = 1, \ldots, n$. Let $\alpha(i,j)$ and $\beta(i,j)$ be two functions:*

$$\alpha(i,j) : \{1, \ldots, n\} \times \{1, \ldots, m\} \to \{1, \ldots, n\};$$
$$\beta(i,j) : \{1, \ldots, n\} \times \{1, \ldots, m\} \to \{1, \ldots, m\}.$$

*The functions $\alpha(i,j)$ and $\beta(i,j)$ specify the desired data movement. Specifically, the data initially stored in the page $p_{i,j}$ are denoted by $D_{i,j}$, and need to be moved into page $p_{\alpha(i,j),\beta(i,j)}$, for all $(i,j) \in \{1, \ldots, n\} \times \{1, \ldots, m\}$.*

*A given number of empty blocks, called auxiliary blocks, can be used in the data movement process, and they need to be erased in the end. To ensure data integrity, at any moment of the data movement process, the data stored in the flash memory blocks should be sufficient for recovering all the original data. The objective is to minimize the total number of block erasures in the data movement process.*

Clearly, the functions $\alpha(i,j)$ and $\beta(i,j)$ together have to form a permutation for the $mn$ pages. To avoid trivial cases, we assume that every block has at least one page whose data need to be moved to another block (otherwise, it can be simply excluded from the set of the $n$ blocks considered in the data movement problem). Also note that a block has to be fully erased whenever any of its pages is modified.

Let us now define some terms that are used throughout the paper. There are two useful graph representations for the data movement problem: the *transition graph* and a *bipartite graph*. In the *transition graph* $G = (V, E)$, $|V| = n$ vertices represent the $n$ data blocks $B_1, \ldots, B_n$. If $y$ pages of data need to be moved from $B_i$ to $B_j$, then there are $y$ directed edges from $B_i$ to $B_j$ in $G$. $G$ is a regular directed graph with $m$ outgoing edges and $m$ incoming edges for every vertex. In the *bipartite graph* $H = (V_1 \cup V_2, E')$, $V_1$ and $V_2$ each has $n$ vertices that represent the $n$ blocks. If $y$ pages of data are moved from $B_i$ to $B_j$, there are $y$ directed edges from vertex $B_i \in V_1$ to vertex $B_j \in V_2$. The two graphs are equivalent but are used in different proofs.

**Definition 2.**
(Block-Permutation Set and Semi-Cycle)
*A set of $n$ pages $\{p_{1,j_1}, p_{2,j_2}, \ldots, p_{n,j_n}\}$ that belong to $n$ different blocks is called a block-permutation set if*

$$\{\alpha(1, j_1), \alpha(2, j_2), \ldots, \alpha(n, j_n)\} = \{1, 2, \ldots, n\}.$$

*If $\{p_{1,j_1}, p_{2,j_2}, \ldots, p_{n,j_n}\}$ is a block-permutation set, then the data they initially store, namely $\{D_{1,j_1}, D_{2,j_2}, \ldots, D_{n,j_n}\}$, are called a block-permutation data set.*

*Let $z \in \{1, 2, \ldots, n\}$. An ordered set of pages*

$$(p_{i_0,j_0}, p_{i_1,j_1}, \ldots, p_{i_{z-1},j_{z-1}})$$

*is called a semi-cycle if for $k = 0, 1, \ldots, z-1$, we have*

$$\alpha(i_k, j_k) = i_{k+1 \bmod z}.$$

**Example 1.** The data movement problem shown in Fig. 1 exemplifies the construction of the transition graph and the bipartite graph. The $nm = 18$ pages can be partitioned into three block-permutation sets:

$$\{p_{1,1}, p_{2,2}, p_{3,2}, p_{4,2}, p_{5,3}, p_{6,1}\},$$
$$\{p_{1,2}, p_{2,1}, p_{3,3}, p_{4,3}, p_{5,2}, p_{6,2}\},$$
$$\{p_{1,3}, p_{2,3}, p_{3,1}, p_{4,1}, p_{5,1}, p_{6,3}\}.$$

The block-permutation sets can be further decomposed into six semi-cycles:

$$(p_{5,3}, p_{1,1}, p_{3,2}, p_{6,1}),$$
$$(p_{2,2}, p_{4,2});$$
$$(p_{5,2}, p_{3,3}, p_{1,2}, p_{2,1}, p_{4,3}, p_{6,2});$$
$$(p_{1,3}),$$
$$(p_{2,3}, p_{3,1}, p_{4,1}),$$
$$(p_{5,1}, p_{6,3}).$$

□

Every *semi-cycle* corresponds to a directed cycle in the transition graph, and every *block-permutation set* corresponds to a set of directed cycles that enter and leave every vertex exactly once. It is not a coincidence that the $nm$ pages in the above example can be partitioned into $m$ block-permutation sets. The following theorem shows it holds for the general case.

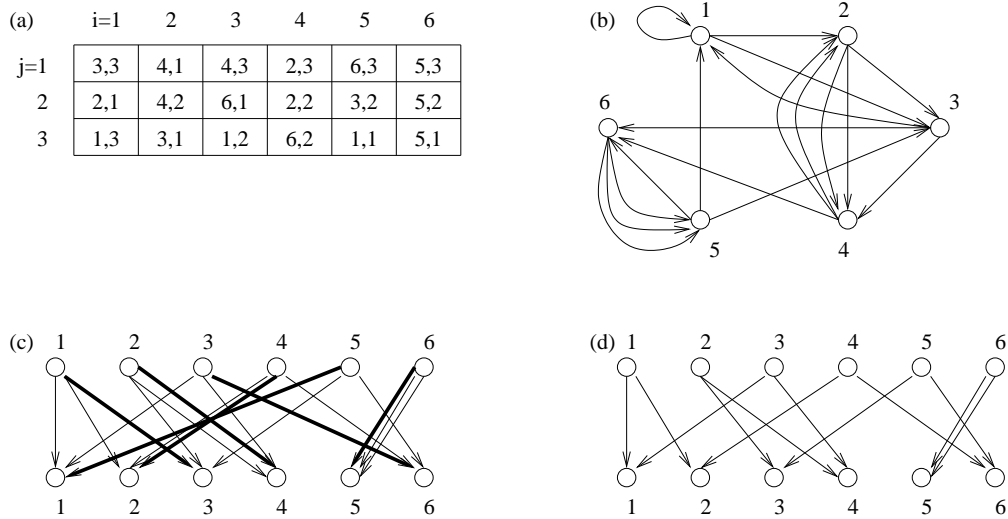

Fig. 1. Data movement with $n = 6, m = 3$. (a) The permutation table. The numbers with coordinates $(i, j)$ are $\alpha(i,j), \beta(i,j)$. For example, $(\alpha(1,1), \beta(1,1)) = (3,3)$, and $(\alpha(1,2), \beta(1,2)) = (2,1)$. (b) The transition graph. (c) The bipartite graph representation. The $n$ thick edges are a perfect matching (a block-permutation set). (d) After removing a perfect matching from the bipartite graph. Here for $i = 1, \ldots, n$, vertex $i$ represents block $B_i$.

**Theorem 1.** *The $nm$ pages can be partitioned into $m$ block-permutation sets. Therefore, the $nm$ pages of data can be partitioned into $m$ block-permutation data sets.*

*Proof:* The data movement problem can be represented by the *bipartite graph*, where every edge represents a page whose data need to be moved into another block. (See Fig. 1 (c) for an example.) It is known that for every bipartite graph $G = (V, E)$ with bipartition $\{A, B\}$ (namely, $A \cap B = \emptyset$ and $A \cup B = V$), we have the Hall's Marriage Theorem [6]:

*The graph $G$ contains a matching of $A$ if and only if $|N(S)| \geq |S|$ for all $S \subseteq A$. (Here $N(S)$ is the set of vertices in the graph $G$ that are adjacent to at least one vertex in $S$. That is, the vertices in $N(S)$ are the neighbors of the vertices in $S$.)*

For the bipartite graph we are considering here, for $i = 1, \ldots, n$, any $i$ vertices in the top layer have $im$ outgoing edges and therefore are connected to at least $i$ vertices in the bottom layer. Therefore, the bipartite graph has a perfect matching. The edges of the perfect matching correspond to a block-permutation set. If we remove those edges, we get a bipartite graph of degree $m - 1$ for every vertex. (See Fig. 1 (c), (d).) With the same argument, we can find another perfect matching and reduce the bipartite graph to regular degree $m - 2$. In this way, we partition the $nm$ edges into $m$ block-permutation sets. ∎

A perfect matching in the bipartite graph can be found using the Ford-Fulkerson Algorithm [14] for computing maximum flow. The idea is to connect all the $n$ top-layer vertices of the bipartite graph to a source $s$ and connect all the $n$ bottom-layer vertices to a sink $t$. Then a perfect matching in the bipartite graph is equivalent to a maximum flow of capacity $n$ between the source $s$ and the sink $t$. The Ford-Fulkerson Algorithm has time complexity $O(n^2 m)$, so decomposing the $nm$ edges in the bipartite graph into $m$ perfect matchings has time complexity $O(n^2 m^2)$. Therefore, we can partition the $nm$ pages into $m$ block-permutation sets in time $O(n^2 m^2)$.

## III. CODING FOR MINIMIZING AUXILIARY BLOCKS

In this paper, we focus on the scenario where as few auxiliary blocks as possible are used in the data movement process. In this section, we show that coding techniques can minimize the number of auxiliary blocks. Afterwards, we will study how to use coding to minimize block erasures.

### A. Data Movement without Coding

When coding is not used, data are directly copied from page to page. The following simple example shows that, in the worst case, more than one auxiliary block is needed for data movement. Note that $D_{i,j}$ denotes the data initially stored in the page $p_{i,j}$.

**Example 2.** Let $n = m = 2$, and let the functions $\alpha(i,j)$ and $\beta(i,j)$ be:
$$(\alpha(1,1), \beta(1,1)) = (1,1),$$
$$(\alpha(1,2), \beta(1,2)) = (2,2),$$
$$(\alpha(2,1), \beta(2,1)) = (2,1),$$
$$(\alpha(2,2), \beta(2,2)) = (1,2).$$

It is simple to verify that without coding, there is no way to move the data as requested with only one auxiliary block. To see that, assume that only one auxiliary block $B_0$ is used. Without loss of generality, assume that we first copy the data in $B_1$ – the data $D_{1,1}$ and $D_{1,2}$ – into $B_0$, and then erase $B_1$. In the next step, the only reasonable choice is to write into $B_1$ the data $D_{1,1}$ and $D_{2,2}$ (which are the data we want to eventually move into $B_1$). After this writing, $B_0$ has $D_{1,1}$ and $D_{1,2}$, $B_1$ has $D_{1,1}$ and $D_{2,2}$, and $B_2$ has $D_{2,1}$ and $D_{2,2}$. The objective of the data movement has not been met yet. However, we can see that there is no way to proceed: in the next step, if we erase $B_0$, the data $D_{1,2}$ will be lost; if we erase $B_2$, the data $D_{2,1}$ will be lost. So the data movement fails. It is simple to verify that no feasible solution exists. Therefore, at least two auxiliary blocks are needed. □

We now show that two auxiliary blocks are sufficient for data movement without coding. The next algorithm uses two auxiliary blocks, which are denoted by $B_0$ and $B_0'$. It operates in a way similar to bubble sort. And it sorts the data of the $m$ block-permutation data sets in parallel.

**Algorithm 1.** (BUBBLE-SORT-BASED DATA MOVEMENT)
*Decompose the $nm$ pages of data into $m$ block-permutation data sets.*

*For $i = 1, \ldots, n-1$*
{
  *For $j = i+1, \ldots, n$*
  {
    *Copy all the data of $B_i$ into $B_0$;*
    *Copy all the data of $B_j$ into $B_0'$;*
    *Erase $B_i$ and $B_j$;*
    *For $k = 1, \ldots, m$*
    {
      *Let $D_{i_1,j_1}$ and $D_{i_2,j_2}$ be the two pages of data in $B_0$ and $B_0'$, respectively, that belong to the k-th block-permutation data set.*
      *Let $p_{i,j_3}$ be the unique page in $B_i$ such that when the data movement process ends, the data stored in $p_{i,j_3}$ will be from the k-th block-permutation data set.*
      *If $\alpha(i_2, j_2) = i$ (which implies $\beta(i_2, j_2) = j_3$ and $\alpha(i_1, j_1) \neq i$), copy the data $D_{i_2,j_2}$ into the page $p_{i,j_3}$; otherwise, copy the data $D_{i_1,j_1}$ into the page $p_{i,j_3}$.*
    }
    *Write into $B_j$ the $m$ pages of data that are in $B_0$ or $B_0'$ but not in $B_i$.*
    *Erase $B_0$ and $B_0'$.*
  }
}

In the above algorithm, for every block-permutation data set, its data are not only sorted in parallel with other block-permutation data sets, but are also always dispersed in $n$ blocks (with every block holding one page of its data). The algorithm uses $O(n^2)$ erasures. (The $n$ blocks $B_1, \ldots, B_n$ are each erased $n-1$ times, while the two auxiliary blocks $B_0$ and $B_0'$ are each erased $\binom{n}{2}$ times.) If instead of bubble sorting, we use more efficient sorting networks such as the Batcher sorting network [2] or the AKS network [1], the number of erasures can be further reduced to $O(n \log^2 n)$ and $O(n \log n)$, respectively. For simplicity we skip the details.

*B. Storage Coding with One Auxiliary Block*

In Algorithm 1, the only function of the two auxiliary blocks $B_0$ and $B_0'$ is to store the data of the data blocks $B_i, B_j$ when the data in $B_i, B_j$ are being swapped. We now show how coding can help reduce the number of auxiliary blocks to one, which is clearly the best possible. Let $B_0$ denote the only auxiliary block, and let $p_{0,1}, p_{0,2}, \ldots, p_{0,m}$ denote its pages. For $k = 1, \ldots, m$, statically store in page $p_{0,k}$ the bit-wise exclusive-OR of the $n$ pages of data in the $k$-th block-permutation data set. We make such a change in Algorithm 1:

*When the data in $B_i, B_j$ are swapped, instead of erasing them together, we first erase $B_i$ and write data into $B_i$, then erase $B_j$ and write data into $B_j$.*

This is feasible because for every block-permutation data set, there are always at least $n$ pages of data related to it stored in the $n + 1$ blocks: $n - 1$ pages of those data are the original data in the block-permutation data set, and the other page of data are the bit-wise exclusive-OR of the data of the block-permutation data set. The total number of block erasures here is of the same order as the algorithm without coding. Therefore, if the AKS network is used for swapping the data, $O(n \log n)$ block erasures will be used in total.

IV. EFFICIENT STORAGE CODING OVER GF(2)

In this section, we present a data movement algorithm that uses only one auxiliary block and $2n$ erasures. It erases every block either once or twice, which is well balanced. The algorithm uses coding over GF(2) and is very efficient.

For convenience, let us assume for now that every block has only one page. The results will be naturally extended to the general case where every block has $m$ pages. Let $B_0$ denote the auxiliary block, and let $p_0$ denote its page. For $i = 1, \ldots, n$, let $p_i$ denote the page in the block $B_i$, and let $D_i$ denote the data initially stored in the page $p_i$. Let

$$\alpha : \{1, \ldots, n\} \to \{1, \ldots, n\}$$

be the permutation such that the data $D_i$ need to be moved into the page $p_{\alpha(i)}$. Let $\alpha^{-1}$ be the inverse permutation of $\alpha$. Say that the $n$ pages $p_1, p_2, \ldots, p_n$ can be partitioned into $t$ semi-cycles, denoted by

$$C_1, C_2, \ldots, C_t.$$

Every semi-cycle $C_i$ ($1 \leq i \leq t$) has a special page called *tail*, defined as follows: if $p_j$ is the *tail* of $C_i$, then for every other page $p_k \in C_i$, we have $j > k$.

We use "$\oplus$" to represent the bit-wise exclusive-OR of data. The following algorithm uses $2n$ block erasures to move data. It consists of two passes: the *forward pass* and the *backward pass*. Note that in the algorithm below, whenever some data are about to be written into a page, the data can be efficiently computed from the existing data in the flash memory blocks (namely, from the data currently stored in the flash memory). The details will be clear later. Also note that for $i = 1, 2, \ldots, n$, $D_{\alpha^{-1}(i)}$ is the data that need to be moved into the page $p_i$.

| $B_0$ | $B_1$ | $B_2$ | $B_3$ | $B_4$ | $B_5$ | $B_6$ | $B_7$ | $B_8$ | Operation |
|---|---|---|---|---|---|---|---|---|---|
| | | | | | forward pass | | | | |
| | **1** | **2** | **3** | **4** | **5** | **6** | **7** | **8** | $\delta_1 \oplus \delta_4$ |
| $1 \oplus 4$ | | 2 | 3 | 4 | 5 | 6 | 7 | 8 | $\delta_2 \oplus \delta_5$ |
| $1 \oplus 4$ | $2 \oplus 5$ | | 3 | 4 | 5 | 6 | 7 | 8 | $\delta_3 \oplus \delta_0 \oplus \delta_4$ |
| $1 \oplus 4$ | $2 \oplus 5$ | $3 \oplus 1$ | | 4 | 5 | 6 | 7 | 8 | $\delta_4 \oplus \delta_7$ |
| $1 \oplus 4$ | $2 \oplus 5$ | $3 \oplus 1$ | $4 \oplus 7$ | | 5 | 6 | 7 | 8 | $\delta_5 \oplus \delta_6$ |
| $1 \oplus 4$ | $2 \oplus 5$ | $3 \oplus 1$ | $4 \oplus 7$ | $5 \oplus 6$ | | 6 | 7 | 8 | copy $\delta_6$ |
| $1 \oplus 4$ | $2 \oplus 5$ | $3 \oplus 1$ | $4 \oplus 7$ | $5 \oplus 6$ | 6 | | 7 | 8 | $\delta_7 \oplus \delta_8$ |
| $1 \oplus 4$ | $2 \oplus 5$ | $3 \oplus 1$ | $4 \oplus 7$ | $5 \oplus 6$ | 6 | $7 \oplus 8$ | | 8 | copy $\delta_8$ |
| $1 \oplus 4$ | $2 \oplus 5$ | $3 \oplus 1$ | $4 \oplus 7$ | $5 \oplus 6$ | 6 | $7 \oplus 8$ | 8 | | |
| | | | | | backward pass | | | | |
| $1 \oplus 4$ | $2 \oplus 5$ | $3 \oplus 1$ | $4 \oplus 7$ | $5 \oplus 6$ | 6 | $7 \oplus 8$ | 8 | | $\delta_7 \oplus \delta_6 \oplus \delta_3 \oplus \delta_0 \oplus \delta_2$ |
| $1 \oplus 4$ | $2 \oplus 5$ | $3 \oplus 1$ | $4 \oplus 7$ | $5 \oplus 6$ | 6 | $7 \oplus 8$ | | 3 | $\delta_6 \oplus \delta_3 \oplus \delta_0 \oplus \delta_2 \oplus \delta_8$ |
| $1 \oplus 4$ | $2 \oplus 5$ | $3 \oplus 1$ | $4 \oplus 7$ | $5 \oplus 6$ | 6 | | 8 | 3 | $\delta_5 \oplus \delta_4 \oplus \delta_1$ |
| $1 \oplus 4$ | $2 \oplus 5$ | $3 \oplus 1$ | $4 \oplus 7$ | $5 \oplus 6$ | | 2 | 8 | 3 | $\delta_4 \oplus \delta_1 \oplus \delta_6$ |
| $1 \oplus 4$ | $2 \oplus 5$ | $3 \oplus 1$ | $4 \oplus 7$ | | 6 | 2 | 8 | 3 | $\delta_3 \oplus \delta_0 \oplus \delta_2 \oplus \delta_8$ |
| $1 \oplus 4$ | $2 \oplus 5$ | $3 \oplus 1$ | | 7 | 6 | 2 | 8 | 3 | $\delta_2 \oplus \delta_8$ |
| $1 \oplus 4$ | $2 \oplus 5$ | | 1 | 7 | 6 | 2 | 8 | 3 | $\delta_1 \oplus \delta_6$ |
| $1 \oplus 4$ | | 5 | 1 | 7 | 6 | 2 | 8 | 3 | $\delta_0 \oplus \delta_3$ |
| | **4** | **5** | **1** | **7** | **6** | **2** | **8** | **3** | |

Fig. 2. Example execution of Algorithm 2. In the rightmost column, $\delta_i$ denotes the data in the page $p_i$ at that moment, for $i = 0, 1, \ldots, 8$.

**Algorithm 2.** (GF(2)-CODING-BASED DATA MOVEMENT)
FORWARD PASS:
    For $i = 1, 2, \ldots, n$ do:
        If $p_i$ is not the tail of its semi-cycle, write

$$D_i \oplus D_{\alpha^{-1}(i)}$$

into the page $p_{i-1}$; otherwise, write

$$D_i$$

into the page $p_{i-1}$. Then, erase the block $B_i$.
BACKWARD PASS:
    For $i = n, n-1, \ldots, 1$ do:
        Write

$$D_{\alpha^{-1}(i)}$$

into the page $p_i$. Then, erase the block $B_{i-1}$.

**Example 3.** Figure 2 gives an example of the execution of Algorithm 2 with $n = 8$ and $t = 2$. Here

$$(\alpha(1), \alpha(2), \ldots, \alpha(8)) = (3, 6, 8, 1, 2, 5, 4, 7).$$

Consequently, we have

$$(\alpha^{-1}(1), \alpha^{-1}(2), \ldots, \alpha^{-1}(8)) = (4, 5, 1, 7, 6, 2, 8, 3).$$

The two semi-cycles are $(p_1, p_3, p_8, p_7, p_4)$ and $(p_2, p_6, p_5)$. In Figure 2, each row is a step of Algorithm 2. The numbers are the data in the blocks. (For convenience, we use $i$ to denote data $D_i$ in the figure for $i = 1, 2, \ldots, 8$.) The rightmost column describes the computation performed for this step, where $\delta_i$ denotes the data in $p_i$ then. □

The correctness of Algorithm 2 depends on whether the data written into a page can always be derived from the existing data in the flash memory blocks. Theorem 2 shows this is true.

**Theorem 2.** *When Algorithm* 2 *is running, at any moment, for $i = 1, 2, \ldots, n$, if the data $D_i$ are not stored in the $n + 1$ blocks $B_0, B_1, \ldots, B_n$, then there must exist a set of data*

$$\{D_i \oplus D_{j_1}, D_{j_1} \oplus D_{j_2}, D_{j_2} \oplus D_{j_3}, \ldots, D_{j_{k-1}} \oplus D_k, D_k\}$$

*that are all stored in the $n + 1$ blocks. Therefore, $D_i$ can be easily obtained by computing the bit-wise exclusive-OR of the data in the set.*

*Proof:* Consider a semi-cycle $C_i$ ($1 \leq i \leq t$), and let us denote its pages by

$$p_{i_1}, p_{i_2}, \ldots, p_{i_x}.$$

Without loss of generality (WLOG), assume

$$\alpha(i_j) = i_{j+1}$$

for $j = 1, 2, \ldots, x - 1$ and

$$\alpha(i_x) = i_1.$$

Also assume that $p_{i_1}$ is the "tail" of the semi-cycle, namely, $i_1 > i_j$ for $j = 2, 3, \ldots, x$. Now imagine a directed path $S$ as follows:
1) $S$ has $x$ vertices, representing the data $D_{i_1}, D_{i_2}, \ldots, D_{i_x}$;
2) There is a directed edge from $D_{i_j}$ to $D_{i_{j+1}}$ for $j = 1, 2, \ldots, x - 1$. The edge represents the data

$$D_{i_j} \oplus D_{i_{j+1}}.$$

For example, the data movement problem in Example 3 has two semi-cycles, $(p_1, p_3, p_8, p_7, p_4)$ and $(p_2, p_6, p_5)$. We show the corresponding directed path $S$ in Fig. 3 (a) and (b).

Consider the *forward pass* in the algorithm. In this pass, for $j = 2, 3, \ldots, x$, right before the data $D_{i_j}$ are erased, the data $D_{i_{j-1}} \oplus D_{i_j}$ are stored. Note that $D_{i_j}$ corresponds to a vertex in the directed path $S$, and $D_{i_{j-1}} \oplus D_{i_j}$ corresponds to the directed edge entering that vertex in $S$. So for every vertex in $S$ whose corresponding data have been erased, there is a directed sub-path in $S$ entering it with this property: "the data represented by the edges in this sub-path, as well as the data represented by the starting vertex of the sub-path, are all stored in the $n + 1$ blocks." This is the same as the condition stated in the theorem. (For instance, for the data movement problem in Example 3, after three block erasures, the stored and un-stored data are as shown in Fig. 3 (c). As an example, consider the erased data $D_3$. The corresponding sub-path entering it contains the data $D_4$, $D_4 \oplus D_1$ and $D_1 \oplus D_3$, which are stored and can be used to recover $D_3$.)

When the *forward pass* of the algorithm ends, the data represented by the vertex $D_{i_1}$ and all the edges in $S$ are all stored in the $n + 1$ blocks. Clearly, all the original data can be recovered.

Now consider the *backward pass* in the algorithm. In this pass, first, the data $D_{i_x}$ are stored and then the data $D_{i_1}$ are erased. Then, for $j = 1, 2, \ldots, x - 1$, right before the data $D_{i_j} \oplus D_{i_{j+1}}$ are erased, the data $D_{i_j}$ are stored. Note that $D_{i_j}$ corresponds to a vertex in the directed path $S$, and $D_{i_j} \oplus D_{i_{j+1}}$ corresponds to the directed edge leaving that vertex in $S$. So for every vertex in $S$ whose corresponding data have been erased, there is a directed sub-path in $S$ leaving it with this property: "the data represented by the edges in this sub-path, as well as the data represented by the end vertex of the sub-path, are all stored in the $n + 1$ blocks." This is the same as the condition stated in the theorem. So the conclusion holds. ■

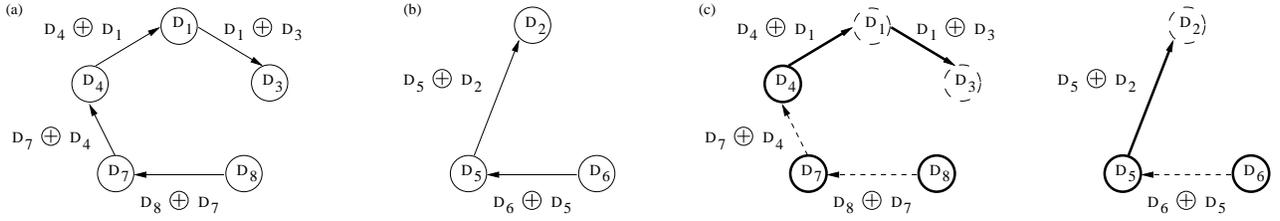

Fig. 3. The directed path $S$ corresponding to a semi-cycle, whose vertices and edges represent data. (a) The directed path $S$ for semi-cycle $(p_1, p_3, p_8, p_7, p_4)$. (b) The directed path $S$ for semi-cycle $(p_2, p_6, p_5)$. (c) The stored and un-stored data after three block erasures in the "forward-pass" of the data-movement algorithm. The vertices and edges of solid thick lines represent the data that are stored at that moment. The vertices and edges of dashed thin lines represent the data that are not stored at that moment.

Algorithm 2 can be easily extended to the general case where every block has $m \geq 1$ pages. Use the algorithm to process the $m$ block-permutation data sets in parallel, in the same way as Algorithm 1. Specifically, for $i = 1, \ldots, n$ and $j = 1, \ldots, m$, let $p_{i,k(i,j)}$ denote the unique page in $B_i$ such that some data in the $j$-th block-permutation data set need to be moved into $p_{i,k(i,j)}$. In the algorithm, every time $B_i$ is erased, write the data related to the $j$-th block-permutation data set into $p_{i,k(i,j)}$. Since every block-permutation set occupies exactly one page in each block, there will be no conflict in writing.

## V. Storage Coding with Minimized Number of Erasures

In this section, we present an algorithm that uses at most $2n - 1$ erasures, which is worst-case optimal. It erases every block either once or twice, which is well balanced. We further show that it is NP hard to minimize the number of erasures for every given instance, but our algorithm provides a 2-approximation.

### A. Optimal Solution with Canonical Labelling

The $n$ blocks initially storing data can be labelled by $B_1, \ldots, B_n$ in $n!$ different ways. Let $y$ be an integer in $\{0, 1, \ldots, n-2\}$. We call a labelling of the $n$ blocks that satisfies the following constraint a *canonical labelling with parameter $y$*:

"For $i = y+1, y+2, \ldots, n-2$ and $j = i+2, i+3, \ldots, n$, no data initially stored in the block $B_j$ need to be moved into the block $B_i$."

Trivially, any labelling is a canonical labelling with parameter $n - 2$. However, given an instance of the data movement problem, it is difficult to find a canonical labelling that minimizes the value of $y$.

We now present a data-movement algorithm for blocks that have a canonical labelling with parameter $y$. It uses one auxiliary block $B_0$, and uses

$$n + y + 1 \leq 2n - 1$$

erasures. (So the smaller $y$ is, the better.) For convenience, let us again assume that every block contains only one page, and let $p_i, D_i, \alpha, \alpha^{-1}$ be as defined in the previous section. Let $r$ denote the number of bits in a page.[1] The algorithm can be naturally extended to the general case, where every block has $m \geq 1$ pages, in the same way as introduced in the previous section.

**Algorithm 3.** (Data Movement with Linear Coding) *This algorithm is for blocks that have a canonical labelling with parameter $y \in \{0, 1, \ldots, n-2\}$. Let $\gamma_1, \gamma_2, \ldots, \gamma_n$ be distinct non-zero elements in the field $GF(2^r)$.*

STEP 1: *For $i = 0, 1, \ldots, y$ do: Erase $B_i$ (for $i = 0$ there is no need to erase $B_0$), and write into $p_i$ the data $\sum_{k=1}^{n} \gamma_k^i D_k$.*

STEP 2: *For $i = y+1, y+2, \ldots, n$ do: Erase $B_i$, and write into $p_i$ the data $D_{\alpha^{-1}(i)}$.*

STEP 3: *For $i = y, y-1, \ldots, 1$ do: Erase $B_i$, and write into the page $p_i$ the data $D_{\alpha^{-1}(i)}$. Finally, erase $B_0$.*

**Theorem 3.** *Algorithm 3 is correct and uses*

$$n + y + 1 \leq 2n - 1$$

*erasures. (Note that the algorithm assumes that the blocks have a canonical labelling with parameter $y$.)*

*Proof:* We show that each time a block $B_i$ is erased, it is feasible to generate all the $n$ pages of original data using the current data stored in the other $n$ pages. Denote by $\delta_i$, $0 \leq i \leq n$, the current data stored in the page $p_i$, which are a linear combination of the $n$ pages of original data. The linear combination written in each page can be represented by a matrix multiplication

$$H \cdot (D_1, D_2, \ldots, D_n)^T = (\delta_0, \ldots, \delta_{i-1}, \delta_{i+1}, \ldots, \delta_n)^T.$$

The matrix $H$ defines the linear combination of the original data written into each page. Consider the first step of the algorithm when the block $B_i$ is erased. The data written in $p_h$, for $0 \leq h \leq i-1$, are

$$\delta_h = \sum_{k=1}^{n} \gamma_k^h D_k,$$

and the data stored in $p_h$, for $i+1 \leq h \leq n$, are

$$\delta_h = D_h.$$

The matrix representation of this problem is

$$\begin{pmatrix} 1 & 1 & \cdots & 1 \\ \gamma_1 & \gamma_2 & \cdots & \gamma_n \\ \gamma_1^2 & \gamma_2^2 & \cdots & \gamma_n^2 \\ \vdots & \vdots & \ddots & \vdots \\ \gamma_1^{i-1} & \gamma_2^{i-1} & \cdots & \gamma_n^{i-1} \\ 0_{(n-i) \times i} & & I_{n-i} & \end{pmatrix} \cdot \begin{pmatrix} D_1 \\ D_2 \\ D_3 \\ \vdots \\ D_{n-1} \\ D_n \end{pmatrix} = \begin{pmatrix} \delta_0 \\ \vdots \\ \delta_{i-1} \\ \delta_{i+1} \\ \vdots \\ \delta_n \end{pmatrix}$$

---

[1]When $r$ is greater than what is needed by Algorithm 3 (which is nearly always true in practice), we can partition each page into bit strings of an appropriate length, and apply the algorithm to the strings in parallel.

where $0_{(n-i) \times i}$ is the zero matrix of size $(n-i) \times i$, and $I_{n-i}$ is the unit matrix of size $(n-i) \times (n-i)$. Since this matrix is invertible, it is feasible to generate all the original data and in particular, the required data that need to be written into $p_i$.

For $i = y+1, y+2, \ldots, n$, after erasing the block $B_i$ during the second step of the algorithm, the data stored in $p_h$, for $0 \le h \le y$, are $\delta_h = \sum_{k=1}^{n} \gamma_k^h D_k$. The data written into $p_h$, for $y+1 \le h \le i-1$, are $\delta_h = D_{\alpha^{-1}(h)}$, and the data stored in $p_h$, for $i+1 \le h \le n$, are $\delta_h = D_h$. These equations are represented as follows:

$$\begin{pmatrix} 1 & 1 & \cdots & 1 \\ \gamma_1 & \gamma_2 & \cdots & \gamma_n \\ \gamma_1^2 & \gamma_2^2 & \cdots & \gamma_n^2 \\ \vdots & \vdots & \ddots & \vdots \\ \gamma_1^y & \gamma_2^y & \cdots & \gamma_n^y \\ & & A_{n-i} & \end{pmatrix} \cdot \begin{pmatrix} D_1 \\ D_2 \\ D_3 \\ \vdots \\ D_{n-1} \\ D_n \end{pmatrix} = \begin{pmatrix} \delta_0 \\ \vdots \\ \delta_{i-1} \\ \delta_{i+1} \\ \vdots \\ \delta_n \end{pmatrix},$$

where $A_{n-i}$ is a matrix of size $(n-y-1) \times n$ defined as follows:

1) The $h$-th row of the matrix $A_{n-i}$ for $1 \le h \le i-y-1$ is a unit vector of length $n$ containing a one in its $(\alpha^{-1}(y+h))$-th entry.
2) The $h$-th row of the matrix $A_{n-i}$ for $i-y \le h \le n-y-1$ is a unit vector that contains a one in its $(y+h+1)$-th entry.

Since there are no data that are moved from block $B_j$ to block $B_i$, where $y+1 \le i \le n-2$ and $i+2 \le j \le n$, the first $i-y-1$ row vectors of the matrix $A_{n-i}$ are different from the last $n-i$ row vectors of the matrix $A_{n-i}$. Therefore, the matrix $A_{n-i}$ contains a set of unit vectors where all the vectors are different from each other. If we calculate the determinant of the matrix on the left hand side according to the rows of the matrix $A_{n-i}$, then we are left with an $(y+1) \times (y+1)$ matrix of the form:

$$\begin{pmatrix} 1 & 1 & 1 & \cdots & 1 & 1 \\ \gamma_{i_1} & \gamma_{i_2} & \gamma_{i_3} & \cdots & \gamma_{i_y} & \gamma_{i_{y+1}} \\ \gamma_{i_1}^2 & \gamma_{i_2}^2 & \gamma_{i_3}^2 & \cdots & \gamma_{i_y}^2 & \gamma_{i_{y+1}}^2 \\ \vdots & \vdots & \vdots & \ddots & \vdots & \vdots \\ \gamma_{i_1}^y & \gamma_{i_2}^y & \gamma_{i_3}^y & \cdots & \gamma_{i_y}^y & \gamma_{i_{y+1}}^y \end{pmatrix}$$

and its determinant is not zero because it is a Vandermonde matrix. Therefore, the matrix on the left hand side is invertible, and it is feasible to generate all the original data $D_i$, $1 \le i \le n$, and in particular the data $D_{\alpha^{-1}(i)}$ that need to be written into the page $p_i$.

For $i = y, y-1, \ldots, 1$, after erasing the block $B_i$ during the third step of the algorithm, the data stored in $p_h$, for $0 \le h \le i-1$, are $\delta_h = \sum_{k=1}^{n} \gamma_k^h D_k$, and the data stored in $p_h$, for $i+1 \le h \le n$, are $\delta_h = D_{\alpha^{-1}(h)}$. Therefore, the matrix representing this equations is

$$\begin{pmatrix} 1 & 1 & \cdots & 1 \\ \gamma_1 & \gamma_2 & \cdots & \gamma_n \\ \gamma_1^2 & \gamma_2^2 & \cdots & \gamma_n^2 \\ \vdots & \vdots & \ddots & \vdots \\ \gamma_1^{i-1} & \gamma_2^{i-1} & \cdots & \gamma_n^{i-1} \\ & & P_{n-i} & \end{pmatrix} \cdot \begin{pmatrix} D_1 \\ D_2 \\ D_3 \\ \vdots \\ D_{n-1} \\ D_n \end{pmatrix} = \begin{pmatrix} \delta_0 \\ \vdots \\ \delta_{i-1} \\ \delta_{i+1} \\ \vdots \\ \delta_n \end{pmatrix},$$

where $P_{n-i}$ is a matrix consisting of $n-i$ row vectors of length $n$, and its $h$-th row vector, $1 \le h \le n-i$, is a unit vector of length $n$ which has a one in its $\alpha^{-1}(i+h)$-th entry and zero elsewhere. As before, all the unit vectors in the matrix $P_{n-i}$ are different from each other. Therefore the matrix on the left hand side is invertible, and it is feasible to generate all the original data $D_i$, $1 \le i \le n$, and in particular the data $D_{\alpha^{-1}(i)}$ that need to be written into the page $p_i$. ∎

The following theorem shows an interesting property of canonical labelling. Note that since every block has some data that need to be moved into it from some other block, every block needs to be erased at least once. So at least $n+1$ erasures (including erasing the auxiliary block) are needed in any case.

**Theorem 4.** *Assume $r$ – the number of bits in a page – is sufficiently large. Let $y \in \{0, 1, \cdots, n-2\}$. There is a data-movement solution using*

$$n+y+1$$

*erasures if and only if there is a canonical labelling of the blocks with parameter $y$.*

*Proof:* First, assume that there is a data-movement solution using $n+y+1$ erasures. Since every block (including the auxiliary block) is erased at least once, there are at least $n-y$ blocks that are erased only once in the solution. Pick $n-y$ blocks erased only once and label them as $B_{y+1}, B_{y+2}, \ldots, B_n$ this way: "in the solution, when $y+1 \le i < j \le n$, $B_i$ is erased before $B_j$." Label the other $y$ blocks as $B_1, \ldots, B_y$ arbitrarily. Let us use contradiction to prove that no data in $B_j$ need to be moved into $B_i$, where $i \ge y+1$, $j \ge i+2$.

Assume some data in $B_j$ need to be moved into $B_i$. After $B_i$ is erased, those data must be written into $B_i$ because $B_i$ is erased only once. When the solution erases $B_{i+1}$ (which happens before $B_j$ is erased), the data mentioned above exist in both $B_i$ and $B_j$. So at this moment, there are at most $nm-1$ pages of distinct data; however, it is impossible to recover all the $nm$ pages of original data using only $nm-1$ pages of distinct data. So there is a contradiction. Therefore, with the above labelling, we have already found a canonical labelling with parameter $y$. The other direction of the proof comes from the existence of Algorithm 3. ∎

We can easily make Algorithm 3 use $2n-1$ erasures by letting $y = n-2$ and using an arbitrary block labelling. On the other hand, $2n-1$ erasures are necessary in the worst case. To see that, consider an instance where $m \ge n$ and every block has some data that need to be moved into every other block. For such an instance, a canonical labelling has to have $y = n-2$, which implies $n+y+1 = 2n-1$ erasures by Theorem 4. So Algorithm 3 is worst-case optimal.

### B. Optimization for Each Given Instance

A specific instance of the data movement problem may require less than $2n-1$ erasures. So it is interesting to find an algorithm that minimizes the number of erasures for each given instance. The following theorem shows that this is NP hard.

**Theorem 5.** *For the data movement problem, it is NP hard to minimize the number of erasures for every given instance.*

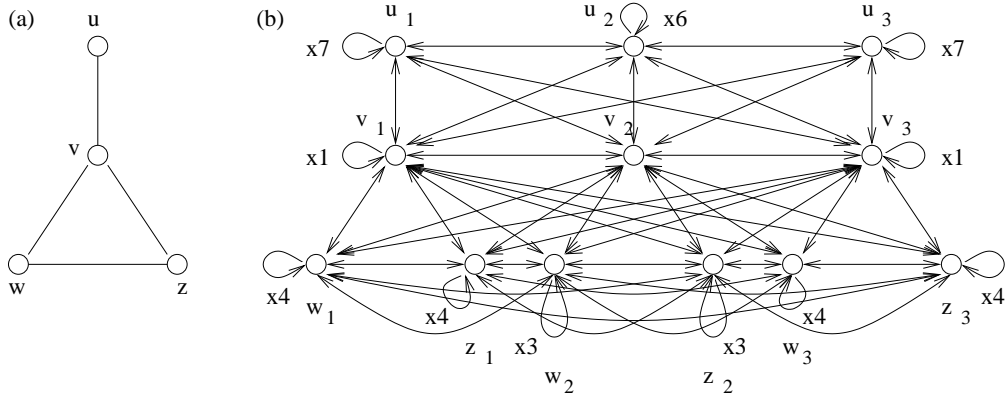

Fig. 4. NP hardness of the data movement problem. (a) A simple undirected graph $G_0$. (b) The corresponding regular directed graph $G'$. Here every edge between two different vertices has arrows on both sides, representing the two directed edges of opposite directions between those two vertices. There is a symbol $\times i$ beside every directed loop, representing $i$ parallel loops of that vertex.

*Proof:* It has been shown in Theorem 4 and its proof that minimizing the number of erasures is as hard as finding a canonical labelling for the blocks with a minimized parameter $y$. So we just need to show that finding a canonical labelling with minimized $y$ is NP hard. We prove it by a reduction from the NP hard MAXIMUM INDEPENDENT SET problem.

Let $G_0 = (V_0, E_0)$ be any simple undirected graph. Let $d(v)$ denote the degree of vertex $v \in V_0$ and let $\Delta = \max_{v \in V_0} d(v)$ denote the maximum degree of $G_0$. We build a regular directed graph $G' = (V_1 \cup V_2 \cup V_3, E')$ as follows. Let $|V_0| = |V_1| = |V_2| = |V_3|$. For all $v \in V_0$, there are three corresponding vertices $v_1 \in V_1, v_2 \in V_2, v_3 \in V_3$. If there is an undirected edge between $u, v \in V_0$ in $G_0$, then there are two directed edges of opposite directions between $u_i$ and $v_j$ for $i = 1, 2, 3$ and $j = 1, 2, 3$. For all $v \in V_0$, there are also two directed edges of opposite directions between $v_1, v_2$ and between $v_2, v_3$. Add some loops to the vertices in $G'$ to make all vertices have the same out-degree and in-degree $3\Delta + 2$. See Fig. 4 for an example.

The graph $G'$ naturally corresponds to a data movement problem with $n = 3|V_0|$ and $m = 3\Delta + 2$, where $G'$ is its *transition graph*. (The transition graph is defined in Section II.) Finding a canonical block labelling with minimized parameter $y$ for this data movement problem is equivalent to finding $t = n - y$ vertices – with the value of $t$ maximized – in $G'$,

$$a_1, a_2, \ldots, a_t,$$

such that for $i = 1, 2, \ldots, t - 2$ and $j = i + 2, i + 3, \ldots, t$, there is no directed edge from $a_j$ to $a_i$. We call such a set of $t$ vertices – with $t$ maximized – the MAXIMUM SEMI-INDEPENDENT SET of $G'$. For all $v \in V_0$, let $N(v)$ denote the neighbors of $v$ in $G_0$.

CLAIM 1: "There is a maximum semi-independent set of $G'$ where $\forall v \in V_0$, either all three corresponding vertices $v_1 \in V_1, v_2 \in V_2, v_3 \in V_3$ are in the set, or none of them is in the set. What is more, if $v_1, v_2, v_3$ are in the set, then no vertex in $\{w_1, w_2, w_3 | w \in N(v)\}$ is in the set."

To prove CLAIM 1, let $(a_1, a_2, \ldots, a_t)$ denote a maximum semi-independent set (*MSS*) of $G'$. (Note that the order of the vertices in the set matters.) Consider two cases:

Case 1: One of $\{v_1, v_2, v_3\}$ is in the MSS of $G'$. WLOG, say it is $v_1$. At most two vertices – say $b$ and $c$ – in $\{w_1, w_2, w_3 | w \in N(v)\}$ can be in the MSS, because otherwise due to the bi-directional edges between them and $v_1$, there would be no way to place them in the MSS. Let us remove $b, c$ from the MSS and add $v_2, v_3$ right after $v_1$ in the MSS. It is simple to see that we get another MSS.

Case 2: Two of $\{v_1, v_2, v_3\}$ are in the MSS of $G'$. WLOG, say they are $v_1$ and $v_2$. At most one vertex – say $b$ – in $\{w_1, w_2, w_3 | w \in N(v)\}$ can be in the MSS, for a similar reason as Case 1. In the MSS, let us remove $b$, move $v_2$ right behind $v_1$, and add $v_3$ right behind $v_2$. Again, we get an MSS.

So in this way, we can easily convert any MSS into an MSS satisfying the conditions in CLAIM 1. So CLAIM 1 is true.

CLAIM 2: "A set of vertices $\{w(1), w(2), \ldots, w(k)\}$ is a maximum independent set of $G_0$ if and only if the set of vertices $(w(1)_1, w(1)_2, w(1)_3, w(2)_1, w(2)_2, w(2)_3, \ldots, w(k)_1, w(k)_2, w(k)_3)$ is an MSS of $G'$." It is simple to see that this is a consequence of CLAIM 1.

So given a canonical labelling with minimized parameter $y$ for the data movement problem with $G'$ as the transition graph, in polynomial time we can convert it into an MSS of $G'$, from that into an MSS of $G'$ satisfying the conditions of CLAIM 1, and finally into a maximum independent set of $G$. So it is NP hard to find a canonical labelling with minimized parameter $y$. So minimizing the number of erasures is NP hard. ∎

Therefore, there is no polynomial time data-movement algorithm that minimizes the number of erasures for every given instance unless $P = NP$. However, since every algorithm uses at least $n + 1$ erasures, and Algorithm 3 can easily achieve $2n - 1$ erasures (by setting $y = n - 2$), we see that the algorithm is a 2-approximation algorithm.

## VI. CONCLUSIONS

In this paper, we study the data movement problem for NAND flash memories. We present sorting-based algorithms that do not utilize coding, which can use as few as $O(n \log n)$ erasures for moving data among $n$ blocks. We show that coding techniques can not only minimize the number of auxiliary blocks, but also reduce the number of erasures to $O(n)$. In particular, we present a solution based on coding over GF(2) that requires only $2n$ erasures. We further present a linear-coding solution that requires at most $2n - 1$ erasures, which is worst-case optimal. Both solutions based on coding achieve an approximation ratio of two with respect to the minimum possible number of block erasures for each instance. They also balance the number of erasures in different blocks very well.

The data movement problem studied here can have numerous practical variations. In one variation, the data to be moved into each block are specified, but the order of the data in that block is allowed to be arbitrary. The algorithms presented in this paper can easily solve this variation of the problem by first assigning an arbitrary page order to each block (which does not affect the performance of the algorithms). In another variation, we are only given a specification as to which group of data needs to be moved into the same block, without specifying which block. Furthermore, the final data may be a function of the data originally stored in the blocks. Such variations require new solutions for optimal performance. They remain open for future research.